\documentclass[prl,superscriptaddress,preprint,showpacs,preprintnumbers]{revtex4}
\usepackage{amssymb}
\usepackage{amsmath}
\usepackage{graphicx}
\usepackage{dcolumn}
\usepackage{bm}
\usepackage{color}

\begin{document}

\title{Spatial wave intensity correlations in quasi-one-dimensional wires}
\author{Gabriel Cwilich}
\affiliation{Department of Physics, Yeshiva University, 500 W 185th Street, New York, NY
10033 USA.}
\author{Luis S. Froufe-P\'erez}
\affiliation{Departamento de F\'{\i}sica de la Materia Condensada and Instituto
``Nicol\'as Cabrera'', Universidad Aut\'{o}noma de Madrid, 28049 Madrid,
Spain}
\author{Juan Jos\'e S\'aenz}
\affiliation{Departamento de F\'{\i}sica de la Materia Condensada and Instituto
``Nicol\'as Cabrera'', Universidad Aut\'{o}noma de Madrid, 28049 Madrid,
Spain}

\begin{abstract}
Spatial intensity correlations between waves transmitted through
random media are analyzed within the framework of the random matrix
theory of transport. Assuming that the statistical distribution of
transfer matrices is isotropic, we found that the spatial
correlation function can be expressed as the sum of three terms,
with distinctive spatial dependences. This result coincides with the
one obtained in the diffusive regime from  perturbative
calculations, but holds all the way from quasi-ballistic transport
to localization. While correlations are positive in the diffusive
regime, we predict a transition to \emph{negative } correlations  as
the length of the system decreases.
\end{abstract}

\date{\today}
 \maketitle

When a wave propagates coherently through a random medium important
correlations emerge between the different propagating paths, which manifest
themselves as correlations in the intensity speckle pattern. They have been
the subject of great interest over the last decade for the case of temporal,
angular, and frequency correlations. \cite%
{Sebbah2,bart-skipe,Stephen,Freund,Feng,MAS,Stone,Bascones,prlcorr}.
Recently, the direct observation of spatial correlations in the intensity
speckle pattern \cite{Sebbah,Emiliani} and in the polarization \cite%
{Chabanov} of electromagnetic waves transmitted through a random
medium has renewed the interest in this problem \cite{Mogollon}.

One of the theoretical approaches followed to study this problem
involved a microscopic diagrammatic calculation
\cite{Feng,Freund,Stephen}. This was also the approach in the recent
work that showed that the spatial correlation function of the
normalized intensity can be expressed as the sum of three terms,
which differ in their spatial dependence \cite{Sebbah}, and in the
work finding an equivalent structure for the correlations in the
polarized radiation \cite{Chabanov}.
While these diagrammatic approaches have the appealing advantage of
illuminating explicitly the nature of the correlations in terms of
the microscopic trajectories of the underlying paths, their
application is strictly limited to the diffusive regime.

An alternative approach, macroscopic in nature, has
been applied successfully to study angular correlations in random media \cite%
{MAS,Stone,Bascones,prlcorr} ; it considers the correlations between
the transport coefficients in the scattering matrix describing the
system in the framework of Random Matrix Theory (RMT)
\cite{Stone,MPK,Bee}. Most of the work based on RMT has been focused
on the study of angular or channel-channel correlations. It shows
that only the assumption of isotropy of the transfer matrix
distribution, discussed below, determines the structure of the
correlations as a function of channel indices. It is the purpose of
this work to apply the RMT approach to study the spatial intensity
correlation functions. Just from the isotropy condition we derive
the structure of the spatial correlations of the normalized
intensity.
Our result, which coincides with the one obtained from microscopic
perturbative calculations (in the diffusive regime)\cite{Sebbah} is
not perturbative and holds all the way from quasi-ballistic
transport to localization. Only the specific values of the
three coefficients ($C_{1}$, $C_{2}$ and $C_{3}$)  depend on
the transport regime. These values can be obtained from the Monte Carlo
solution of the Dorokhov, Mello, Pereyra, and Kumar
(DMPK)\cite{MPK,Dorokhov} scaling equation, and are in full agreement
with microscopic numerical calculations of bulk disordered wires.
While the correlations are positive in the diffusive regime, we predict
a transition to \emph{negative } values for both angular and
spatial correlations as the length of the system decreases.


We will consider a wave propagating in the $z$-direction in a
constrained geometry (see the inset of Figure 1). The input and
output faces of the disordered region are the planes $z=0$ and $z=L$
respectively. The transverse coordinates in the system are described
by $\roarrow{\rho}$. \ We will only discuss the case of scalar waves
in this work, neglecting polarization effects. \ The eigenfunctions
of the cavity (in the absence of disorder) naturally separate into a
longitudinal and a transverse part,
\begin{equation}
\phi _{n}^{\pm }(\roarrow{r})=\frac{1}{\sqrt{k_{n}}}\psi _{n}(\roarrow{\rho}%
)\exp \left\{\pm ik_{n}z\right\}
\end{equation}%
The integer $n=1,2,\dots ,N$ labels the propagating modes, also referred to
as scattering channels. Mode $n$ has a real wave number $k_{n}=\sqrt{%
k^{2}-q_{n}^{2}}$ , where $k$ is the wavenumber of the incident
radiation, and $q_{n}$ is the momentum associated with the
normalized transverse wave function $\psi _{n}(\roarrow{\rho})$. The
normalization of the total wave function $\phi _{n}$ is chosen to
carry unit current.
%
The scattering matrix $\mathbf{S}$
\begin{equation}
\mathbf{S} = \left(
\begin{array}{cc}
\mathbf{r} & \tilde{\mathbf{t}} \\
\mathbf{t} & \tilde{\mathbf{r}}%
\end{array}%
\right) .
\end{equation}%
relates the asymptotic propagating outgoing waves to the incoming
ones:
Matrix elements $r_{ba}$ and $t_{ja}$ denote the reflected amplitude
in channel $b$ and the transmitted amplitude in channel $j$ when
there is a unit flux incident from the left in channel $a$;
$\tilde{r}_{ji}$ and $\tilde{t}_{bi}$ have an analogous meaning when
the incident flux in channel $i$ comes from the right. Flux
conservation and reciprocity imply that the matrix $\mathbf{S}$ is
unitary and symmetric.


Let us  consider two point sources at $\roarrow{r}=%
\roarrow{r_A}$  and \newline $\roarrow{r}=%
\roarrow{r_B}$ on the left hand side of the system, and two detectors at $\roarrow{r}=%
\roarrow{r_1}$ and $\roarrow{r}=%
\roarrow{r_2}$ on the right side (as sketched in Fig. 1).
For a point source at $\roarrow{r}=%
\roarrow{r_A}$ the incoming field from the left is proportional to
the Green function of the clean waveguide,
\begin{equation}
G_{0}^{+}(\roarrow{r}_{A},\roarrow{r})=\frac{i}{2}\sum_{a}\phi _{a}^{+\ast }(%
\roarrow{r}_{A})\phi _{a}^{+}(\roarrow{r})\ \ \ ;  \ \ (z>z_{A}),
\label{Go}
\end{equation}%
The field at a point $%
\roarrow{r}_{1}$ on the right side outside the system will be given
by
\begin{equation}
E(A,1)=\sum_{j} \left\{ \sum_{a} t_{ja} c_{a} \right\} \phi
_{j}^{+}(\roarrow{r_1}) \label{defo}
\end{equation}%
where $c_{a} \propto \phi _{a}^{+\ast}(\roarrow{r}_{A})$. The
average intensity at that point is given by  $\langle I(A,1)\rangle
\equiv \langle |E(A,1)|^{2}\rangle$, where $<...>$ denotes disorder
averaging (over the ensemble of samples). For a finite sample, the
transmission amplitudes, $t_{ja}$, are assumed to have random phases
with
$\left\langle t_{ja}\right\rangle = 0$ and 
$\left\langle t_{ja}t_{j^{\prime }a^{\prime }}^{\ast }\right\rangle
= \left\langle T_{ja}\right\rangle \delta_{aa^{\prime
}}\delta_{jj^{\prime }}$.
The square of the field-field spatial correlation function can be
written as
\begin{eqnarray}
C_1(A,1;B,2) &=& \frac{|\langle E(A,1) E^{\ast }(B,2)
\rangle|^2}{\langle I(A,1)\rangle \langle I(B,2)\rangle}
\label{defffcf}
\end{eqnarray}%
where
\begin{eqnarray}
\langle E(A,1) E^{\ast }(B,2) \rangle &=& \sum_{aa^{\prime }
jj^{\prime }}c_{a}d_{a^{\prime }}^{\ast }\phi
_{j}^{+}(\roarrow{r_1})\phi _{j^{\prime }}^{+\ast
}(\roarrow{r_2})\left\langle
t_{ja}t_{j^{\prime }a^{\prime }}^{\ast }\right\rangle \nonumber \\
&=& \sum_{a j}c_{a} d_{a}^{\ast } \phi _{j}^{+}(\roarrow{r_1})\phi
_{j}^{+\ast }(\roarrow{r_2})\left\langle T_{ja} \right\rangle.
\label{defffcf1}
\end{eqnarray}%
(with $d_{a} \propto \phi _{a}^{+}(\roarrow{r}_{B})$).
 The square
of the field-field correlation function takes a simple form in the
equivalent channel approximation: Assuming that $\left\langle T_{ja}
\right\rangle = (1/N^2)\sum_{ja} \left\langle T_{ja} \right\rangle
\equiv g/N^2$, equation (\ref{defffcf1}) factorizes and
\begin{equation}
C_1(A,1;B,2) = |F(\roarrow{r}_{A},\roarrow{r}_{B})|^{2}
|F(\roarrow{r}_{1},\roarrow{r}_{2})|^{2}
\end{equation}
where $|F(1,2)|^{2}$ can be written in terms of the Green function
(\ref{Go}):
\begin{equation}
|F(\roarrow{r}_{1},\roarrow{r}_{2})|^{2}=\frac{|\Im \{G_{0}^{+}(\roarrow{r}%
_{1},\roarrow{r}_{2})\}|^{2}}{\Im \{G_{0}^{+}(\roarrow{r}_{1},\roarrow{r}%
_{1})\}\Im \{G_{0}^{+}(\roarrow{r}_{2},\roarrow{r}_{2})\}}.
\label{la_g}
\end{equation}
In the large-$N$ limit, the Green function of the clean waveguide
tends to the free-space Green function, $\exp (ikr)/(4\pi r)$ in the
case of a 3D conductor, and we have
\begin{equation}
|F(\roarrow{r}_{1},\roarrow{r}_{2})|^{2}\approx \left\vert
\frac{\sin (k|\Delta \roarrow{r}_{12}|)}{k|\Delta
\roarrow{r}_{12}|}\right\vert ^{2} \label{sinc}
\end{equation}%
and analogous results holds in 2D with the ``sinc'' replaced by the
Bessel function $J_{0}$:
\begin{equation}
|F(\roarrow{r}_{1},\roarrow{r}_{2})|^{2}\approx J_{0}^{2}(k|\Delta %
\roarrow{r}_{12}|)  \label{bessel}.
\end{equation}
The behavior of the function $|F|^2$ versus $k\Delta r = k|x_1-x_2|$
is illustrated in Figure 1. Black dots correspond to the large-$N$
limit, eq. (\ref{bessel}).
For a finite width, the field correlation function (eq. \ref{la_g})
strongly depends on the position of the detectors. Finite size
effects can be minimized by using the experimental approach of ref.
\cite{Sebbah}: we consider the average of $|F|^2$ when $x_1$ and
$x_2$  are uniformly distributed over the interval $(W/10,W-W/10)$.
As can be seen in Figure 1, the averaged $|F|^2$ (dashed line) already presents the typical
large-$N$ behavior for the case of $N=20$.

\begin{figure}
\includegraphics[width=8cm]{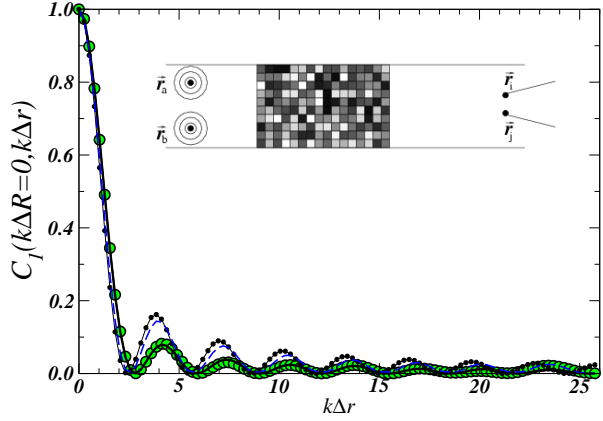}
\caption{Square of the field correlation function for a single
source as a function of the distance between detectors for a 2D
waveguide. The black dots correspond to the expected large-$N$ limit
(eq. \ref{bessel}) assuming that all channels are equivalent. For
finite $N$ ($N=20$), the results represent the average over
different source and detector positions: The large circles are the results of
microscopic numerical calculations (for $g = 1.12$) based on the
model system
sketched in the inset.
The continuous line is the result of equations  \ref{defffcf} and
\ref{defffcf1} with $\langle T_{aj}\rangle $ obtained from the
numerical calculations while the dashed line corresponds to the
equivalent channel approach, $\langle T_{aj}\rangle = g/N^2$ (eq.
\ref{la_g}). 
}
\end{figure}

The spatial intensity correlation function can be defined as
\begin{equation}
C(A,1;B,2)\equiv \langle I(A,1)I(B,2)\rangle -\langle I(A,1)\rangle
\langle I(B,2)\rangle \end{equation} where the first term of the
right hand side is given by
\begin{multline}
\langle I(A,1)I(B,2)\rangle =\sum_{aa^{\prime }bb^{\prime
}}\sum_{ii^{\prime }jj^{\prime }}\left\{ \left( c_{a}c_{a^{\prime
}}^{\ast }d_{b}d_{b^{\prime
}}^{\ast }\right) \right. \times  \label{def2} \\
\left. \left( \phi _{j}^{+}(\roarrow{r}_{1})\phi _{j^{\prime }}^{+\ast }(%
\roarrow{r}_{1})\phi _{i}^{+}(\roarrow{r}_{2})\phi _{i^{\prime }}^{+\ast }(%
\roarrow{r}_{2})\right) \left\langle t_{ja}t_{j^{\prime }a^{\prime
}}^{\ast }t_{ib}t_{i^{\prime }b^{\prime }}^{\ast }\right\rangle
\right\}
\end{multline}%
%
\begin{figure}
\includegraphics[width=8cm,angle=0]{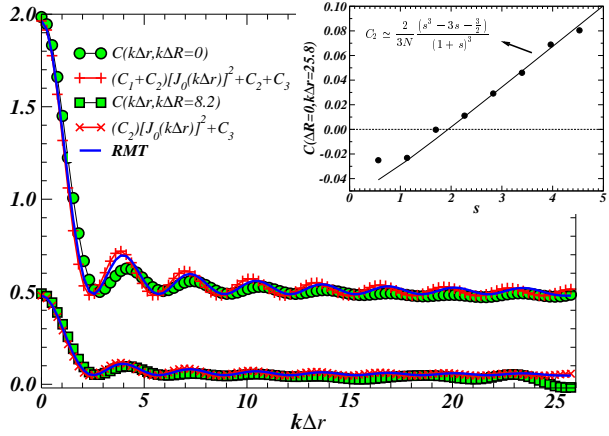}
\caption{ Plots of the intensity correlation function. Inset: Long
range intensity correlations ($C_2$) as a function of the length
$s=L/\ell$ of the system. Black dots are the results of numerical
calculations.
 }
\end{figure}

In contrast with field-field  correlations, the calculation of the
averages presents subtle properties directly related to the symmetry
and statistical properties of the $\mathbf{S}$ matrix.
Calling $%
{\mathcal{T}}_{1},{\mathcal{T}}_{2},\dots ,{\mathcal{T}}_{N}$ \ the
set of \
N common transmission eigenvalues of the four Hermitian matrices $\mathbf{{tt%
}^{\dagger }}$, $\tilde{\mathbf{t}}\tilde{\mathbf{t}}^{\dagger }$, $\mathbf{{%
rr}^{\dagger }}$ and $\tilde{\mathbf{r}}\tilde{\mathbf{r}}^{\dagger
}$, the matrix $S$ can also be written in terms of the
${\mathcal{T}}_{n}\ $'s by means of the polar decomposition
\cite{MPK,Bee}
\begin{equation*}
\mathbf{S}=\left(
\begin{array}{cc}
\mathbf{u}^{(1)} & 0 \\
0 & \mathbf{u}^{(2)}%
\end{array}%
\right) \left(
\begin{array}{cc}
-\sqrt{1-\mathcal{\tau }} & \sqrt{\mathcal{\tau }} \\
\sqrt{\mathcal{\tau }} & \sqrt{1-\mathcal{\tau }}%
\end{array}%
\right) \left(
\begin{array}{cc}
\mathbf{u}^{(1)\top} & 0 \\
0 & \mathbf{u}^{(2)\top}%
\end{array}%
\right)
\end{equation*}%
where $\mathbf{u}^{(i)}$ are unitary matrices and \newline ${\mathcal{T}}=\text{diag}(%
{\mathcal{T}}_{1},{\mathcal{T}}_{2},\dots ,{\mathcal{T}}_{N})$ is a
$N\times N$ diagonal matrix with the transmission eigenvalues on the
diagonal.
The
transmission and reflection matrices can, then, be written as
\begin{eqnarray}
r_{ba} &=&-\sum_{n}u_{bn}^{(1)}\left(
\sqrt{1-{\mathcal{T}}_{n}}\right)
u_{an}^{(1)} \\
t_{ja} &=&\sum_{n}u_{jn}^{(2)}\left( \sqrt{{\mathcal{T}}_{n}}\right)
u_{an}^{(1)}
\end{eqnarray}

One of the key assumptions in the macroscopic approach is the
hypothesis of \emph{isotropy} \cite{MPK,Bee,Stone} . Under this
hypothesis the statistical distribution of the transmission
eigenvalues $\{{\mathcal{T}}_{n}\}$ is independent of the unitary
matrices $\mathbf{u}^{(i)}$, and the calculation
of the statistical averages in (\ref{def2}) factorizes. \ Moreover, $\mathbf{%
u}^{(1)}$ and $\mathbf{u}^{(2)}$ are statistically independent from each
other, each being distributed according the invariant measure of the unitary
group.
By using the averages over the unitary group $\langle (u_{jn})(u_{j^{\prime
}n^{\prime }})^{\ast }\rangle $ and $\langle (u_{jn}u_{im})(u_{j^{\prime
}n^{\prime }}u_{i^{\prime }m^{\prime }})^{\ast }\rangle $ (evaluated by
Mello in ref. \cite{Mello}), after some algebra,
we find
\begin{equation}
\langle t_{ja}t_{j^{\prime }a^{\prime }}^{\ast }\rangle =\frac{1}{N^{2}}%
\langle T\rangle \delta _{jj^{\prime }}\delta _{aa^{\prime }}  \label{tt}
\end{equation}%
\begin{multline}
\langle t_{ja}t_{j^{\prime }a^{\prime }}^{\ast }t_{ib}t_{i^{\prime
}b^{\prime }}^{\ast }\rangle = \\
\left[ A_{N}\langle T^{2}\rangle -B_{N}\langle T_{2}\rangle \right] \left(
\delta _{ij^{\prime }}\delta _{i^{\prime }j}\delta _{ab^{\prime }}\delta
_{a^{\prime }b}+\delta _{jj^{\prime }}\delta _{ii^{\prime }}\delta
_{aa^{\prime }}\delta _{bb^{\prime }}\right)  \\
+\left[ A_{N}\langle T_{2}\rangle -B_{N}\langle T^{2}\rangle \right] \left(
\delta _{jj^{\prime }}\delta _{ii^{\prime }}\delta _{ab^{\prime }}\delta
_{a^{\prime }b}+\delta _{ij^{\prime }}\delta _{i^{\prime }j}\delta
_{aa^{\prime }}\delta _{bb^{\prime }}\right)   \label{ttt}
\end{multline}%
introducing the notation
$g = \langle T\rangle \equiv \sum_{n}\langle
{\mathcal{T}}_{n}\rangle$, \newline $\text{var}\{g\}\equiv \langle
T^{2}\rangle -\langle T\rangle ^{2}$,
$\langle T_{2}\rangle  \equiv \sum_{n}\langle
{\mathcal{T}}_{n}^{2}\rangle$
and the coefficients $\ A_{N}$ \ and $B_{N}$ given by
\begin{equation}
A_{N}=\frac{N^{2}+1}{N^{2}}\frac{1}{(N^{2}-1)^{2}}\quad ;\quad B_{N}=\frac{2%
}{N}\frac{1}{(N^{2}-1)^{2}}  \notag
\end{equation}

>From equation (\ref{ttt}), and taking $i=i^{\prime },j=j^{\prime
}$,\newline $a=a^{\prime },b=b^{\prime }$, we easily recover the well known
channel-channel correlation function $C_{jaib}$ \cite{MAS}
\begin{multline}
\frac{\langle T_{ja}T_{ib}\rangle }{\langle T_{ja}\rangle \langle
T_{ib}\rangle }-1= C_1 \left( \delta _{ij}\delta _{ab}\right) + C_2 \left(
\delta _{ab}+\delta _{ij}\right) + C_3  \label{angular}
\end{multline}
where
\begin{eqnarray}
C_1 &\equiv& \frac{N^{4}}{\langle T\rangle ^{2}}\left[ A_{N}\langle
T^{2}\rangle -B_{N}\langle T_{2}\rangle \right] \\
C_2 &\equiv& \frac{N^{4}}{\langle T\rangle ^{2}}\left[ A_{N}\langle
T_{2}\rangle -B_{N}\langle T^{2}\rangle \right] \\
C_3 &\equiv& C_1 -1
\end{eqnarray}
The angular correlation function has the same structure as that found first
by Feng \emph{et al.} \cite{Feng}, the three coefficients $C_1$, $C_2$ and $%
C_3$ corresponding, respectively, to short, long and infinite range
correlations.

In order to obtain the spatial correlation function, we substitute
the expressions (\ref{tt}) and (\ref{ttt}) for the averages in
equations (\ref{defffcf1}) and (\ref{def2}),
\begin{eqnarray}
&&
C \left (\Delta \roarrow{r}_{ab},\Delta \roarrow{r}_{12}\right )
\equiv
\frac{\langle I(a,1)I(b,2)\rangle }{\langle I(a,1)\rangle \langle
I(b,2)\rangle }-1=  \notag \\
&& C_{1} \left( \left\vert F(\roarrow{r}_{a},\roarrow{r}_{b})\right\vert
^{2}\left\vert F(\roarrow{r}_{1},\roarrow{r}_{2})\right\vert
^{2}\right)
\notag \\
+ && C_{2} \left( \left\vert F(\roarrow{r}_{a},\roarrow{r}_{b})\right\vert
^{2}+\left\vert F(\roarrow{r}_{1},\roarrow{r}_{2})\right\vert
^{2}\right)  + C_{3}  \label{final}
\end{eqnarray}%
where $|F(1,2)|^{2}$ is the square of the field correlation function (eq. \ref{la_g})
and the correlation coefficients $C_{1}$, $C_{2}$ and $C_{3}$ are \emph{%
exactly the same} as those appearing in equation (\ref{angular}).
The structure of the spatial correlations is, then, equivalent to
that obtained for channel correlations with the angular
\textquotedblleft $\delta
_{ab}$\textquotedblright\ functions replaced by the spatial functions $%
|F(a,b)|^{2}$.

The behavior of the spatial correlations is illustrated in Figure 2.
For a single source and two detectors ($\Delta \roarrow{r}_{ab} =
\Delta R = 0 $ and   $\Delta \roarrow{r}_{12}\equiv \Delta r$), or
viceversa, the correlation function (in a 2D waveguide) is given by
$\approx (C_1+C_2) (J_0(k \Delta r))^2 + C_2 + C_3$ and approach a
constant as $\Delta r$ increases. For large $\Delta R$, the
correlations as a function of $\Delta r$ behave as $\approx C_2
(J_0(k \Delta r))^2 + C_3$. These results are in qualitative
agreement with the results of reference \cite{Sebbah}. As a matter
of fact, our
expressions (\ref{final}) and (\ref{la_g}) ( or %
\ref{sinc} in the large-$N$ limit) are consistent, after a slight
reordering of the terms, with the expression (5) in reference
\cite{Sebbah}, which was obtained through a perturbative
diagrammatic expansion. The equivalence with polarization
correlations (equation (2) of reference \cite{Chabanov}) is also
evident. However, while the diagrammatic expansions are strictly
valid in the diffusive regime, our result \textbf{does not depend at
all on the transport regime}, and are a direct consequence of the
isotropy hypothesis. Only the relative \emph{size} of $C_{1}$,
$C_{2}$ and $C_{3}$ will depend on the length of the system $L$ and
the mean free path $\ell $ through the
distribution of transmission eigenvalues $P(\{{\mathcal{T}}_{n}\},s)$, with $%
s=L/\ell $. Figure 3 summarizes the dependence of the correlation
coefficients with the length of the system as obtained from a Monte
Carlo calculations of the DMPK equation \cite{prl_luis}.
%
In the diffusive regime ($1\ll g\ll N$), the correlation factors
take the
well known values \cite{Feng,MAS}: $C_{1}\approx 1$, $gC_{2}\approx 2/3$ and $%
g^{2}C_{3}\approx 2/15$. As the length of the system decreases the
exact solution of the DMPK equation is well described by the known
analytical results based on $1/N$ expansion \cite{prlcorr} (dashed
line in Fig. 3).
While the correlations are positive in the diffusive regime, our
results predict \emph{a transition to negative correlations for both
the angular and the spatial case} (see inset in Fig. 2).

\begin{figure}
\includegraphics[height=8cm,angle=-90]{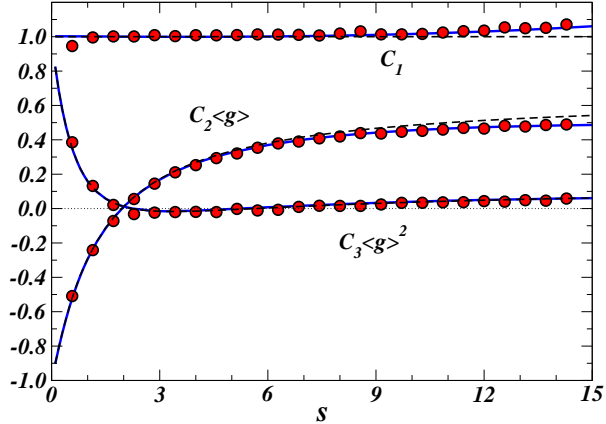}
\caption{ Correlation coefficients $C_1$, $C_2$ and $C_3$ as a
function of the length $s=L/\ell$ of the system. Continuous lines
are the results of the DMPK equation. Dashed line represents the
perturbative $1/N$ expansion (after ref. \cite{prlcorr}). Circles
are numerical results for the model system sketched in Fig. 1 }
\end{figure}

In order to confirm the predictions of the macroscopic approach, we have
performed extensive numerical calculations based on the simple
two-dimensional (2D) model  sketched in Fig. 1.  For electromagnetic waves,
we assume s-polarization with the electric vector parallel to the walls. The
disordered region is divided in small rectangular regions of section $\delta
_{z}\times \delta _{x}$. Within each slice the refraction index $n_{R}$ has
random values distributed uniformly in the interval $(1-\delta
n_{R},1+\delta n_{R})$. We  take $\delta _{x}=W/10$, $\delta _{z}=W/20$ and $%
\delta n_{R}=0.025,$ for a 20-mode waveguide ($ W/\lambda =10.25$)
as in Figure 1. Transmission and reflection coefficients are exactly
calculated by solving the 2D wave equation by mode matching at each
$\delta _{z}$-slice, together with a generalized scattering-matrix
technique \cite{Torres04}. Figure 1 shows the behavior of the
averaged field correlations (for $s=15$, $\langle g \rangle=1.12$).
The numerical results show that transport is not fully isotropic
(the distribution of the transmission coefficients $T_{ja}$ has some
dependence on the particular values of $j$ and $a$). The numerical
intensity correlations are very close to the expected behavior (see
Fig. 2) although  they  are not fully described by equations
\ref{angular} and \ref{final}. However we can extract the
correlation coefficients $C_{1},C_{2}$ and $C_{3}$  from a least
square fitting of  spatial or angular numerical correlations with
equation (\ref{final}) or (\ref{angular}) respectively. In both cases, the
obtained coefficients  (circles in Figure 3) are in full agreement
with the predictions of the DMPK approach.


%
%
%

 This work has
been supported by the Spanish MCyT (Ref. No. BFM2003-01167) and the
European Integrated Project "Molecular Imaging"
(LSHG-CT-2003-503259).
One of us (GC) wants to acknowledge the hospitality of the D.F.M.C. (U.A.M. - Spain) where this work was completed.

\bigskip

\bigskip

\end{document}